\documentclass[11pt]{article}
\usepackage[a4paper,hmarginratio=1:1,vmarginratio=2:3,totalwidth=15.9cm,
  totalheight=21.5cm]{geometry}  
\usepackage{latexsym}
\renewcommand{\baselinestretch}{1.38}   
\usepackage{epsfig}

\newcommand{\cG}{{\cal G}}

\newcommand{\cJ}{{\cal J}}

\newcommand{\cL}{{\cal L}}

\newcommand{\cO}{{\cal O}}

\newcommand{\cZ}{{\cal Z}}

\newcommand{\bZ}{{\bf Z}}

\newcommand{\ra}{\rightarrow}
\newcommand{\be}{\begin{equation}}
\newcommand{\ee}{\end{equation}}
\newcommand{\bea}{\begin{eqnarray}}
\newcommand{\eea}{\end{eqnarray}}

\DeclareMathSymbol{\mg}{\mathrel}{symbols}{"1D}

\newcounter{oldcounter} 
\addtocounter{equation}{1}
\begin{document} 
\begin{flushright}  
%{OUTP-0620P}\\
\end{flushright}  
\vskip 3 cm
\begin{center} 
{\Large{\bf Threshold Effects  near
 the Compactification Scale\\
 for Gauge Couplings on Orbifolds\\}}

\vspace{1.cm} 
{\bf D.~M.~Ghilencea\footnote{E-mail Address: d.ghilencea1@physics.ox.ac.uk}}\\
\vspace{0.8cm} 
{\it  Rudolf Peierls Centre for Theoretical Physics,
University of Oxford, \\
1 Keble Road, Oxford, OX1 3NP, United Kingdom.}\\
\end{center} 
\vspace{0.5cm}
\begin{abstract}
\noindent
In 6D orbifold compactifications of N=1 supersymmetric gauge theories,
the one-loop behaviour of the  4D  {\it effective} gauge coupling and 
of its beta function are carefully investigated for 
momentum scales $k^2$  near  the compactification scale(s).
It is showed that as $k^2$ is crossing the compactification scale(s), there 
exists a smooth transition (``crossover'') to a power-like behaviour 
of the 4D effective coupling,  with a continuous  one-loop beta function.
Contrary to the naive expectation, 
the power-like behaviour  sets in at momentum scales 
{\it smaller} than the compactification  scale, to suggest a
 {\it global} effect of the Kaluza-Klein modes even below this scale.
We argue that the experimental observation of such  behaviour  is not 
necessarily a quantum effect of the compactification,
as often thought, and is  dominated by  classical effects.
 Special attention is paid to convergence issues near
compactification scale(s) and to the scheme dependence of the results. 
\end{abstract}

\thispagestyle{empty}
\newpage

\section{Introduction}

The study of physics beyond the Standard Model (SM) considered 
in the past the quantum corrections that compactification brings
to the SM  gauge couplings, when the SM or its supersymmetric
version are embedded in a higher dimensional theory.
Such corrections are important in models  studying gauge
 unification, but also in the context of ``large'' extra dimensions,
when  changes to the 4D logarithmic
running of the couplings could be observed experimentally.
There is a large amount of work on this topic in 
effective field theory approaches 
or string models \cite{RC1}-\cite{FW}.

In the context of  4D N=1 supersymmetric models obtained 
from  an orbifold compactification of a higher dimensional gauge theory,
one particular problem  was however  little investigated. 
This concerns the behaviour of the associated quantum corrections to the 
4D effective gauge coupling for {\it momentum scales}
{\it near}  the compactification scale(s). 
This paper will investigate in detail this behaviour
for 4D N=1 field theory  models obtained from a 6D N=1 supersymmetric 
effective gauge theory compactified on $T^2/Z_2$.
Another goal is to re-consider
 the  widespread opinion in phenomenological studies that a
power-like running of the 4D effective gauge coupling is 
a proof of the existence of ``large'' extra dimensions.

To achieve these goals string theory is, rather unfortunately,
 of little help. This is because string theory  computes only
{\it on-shell} values of the 2-point Green function of the gauge boson
 self-energy. Therefore it cannot describe the {\it running} of
the 4D effective gauge  coupling with respect to the momentum scale, and
in particular near the compactification scale(s). 
An off-shell effective field theory
approach, although it may miss some non-perturbative
effects,  can consistently  address this problem.  
It turns out that the one-loop running of the 4D effective gauge
couplings in  gauge theories on orbifolds is controlled
by higher dimensional operators \cite{nibbelink,DHK,santamaria}, with
interesting results. Such effects  are not addressed 
by string theory.

In \cite{DHK} an off-shell calculation of the gauge
boson self energy was performed at one-loop level in 6D N=1 supersymmetric
gauge theories on $T^2/Z_2$  in both Abelian and non-Abelian cases.
The method employed  an orbifold-compatible background field method
to compute the one-loop correction to the 4D effective gauge coupling
and its dependence on the momentum scale.
 It was found that bulk corrections, when integrated out, generated
 bulk higher dimensional (derivative)
 operators as counterterms to the gauge couplings running, in addition
to brane localised gauge kinetic terms. While the latter ultimately generate 
only  a logarithmic correction in the momentum
to the 4D effective gauge coupling, 
the higher derivative operator also generates terms proportional
to $k^2 v/h^2_{\rm tree}$, where $v$
 is the volume of compactification,
$k^2$ the external momentum scale and $h_{\rm tree}$ is the coupling of the
higher derivative operator. For $k^2\ll 1/v$  the higher derivative term
is strongly suppressed,  the usual 4D N=1 logarithmic running is
present and the theory appears as 4D renormalisable.
 As we increase the momentum scale, the higher derivative
operator becomes more important and eventually dominates the  running
of the coupling for $k^2$ larger than the compactification scale(s).
This is the so-called ``power-like'' running region.
 However, near the compactification scale, when $k^2\sim 1/v$,
 the analytical formulae provided in \cite{DHK}  are very complicated and less
illuminating, and this region was not  studied in the past\footnote{Related
  problems are discussed in  \cite{CQ} for a scalar theory in 6D.}.
The  purpose of this work is to use a numerical approach to
 explore in  detail the  effects of ``opening up''
of extra-dimensions near the compactification scale and the  running of
 the 4D effective gauge coupling  {\it across} this threshold.
 Special attention is paid to 
convergence issues in this region.
We are  not aware of a similar, prior study of this kind.

A problem that affects such analysis is that in 
any 4D effective theory one can   in principle include
in the  running of the 4D gauge coupling, the effect 
of a tree-level higher dimensional (derivative) operator
which can be present in the effective action.
 This operator can have a form similar
to that generated dynamically by compactification,
and can be present at classical level  with a  coefficient
suppressed by some  high scale (rather than by the compactification
scale(s)). The question is then if one can distinguish such ``classical'' 
corrections to the gauge coupling from those 
when such operator is generated at the quantum level, in
 an orbifold compactification of type discussed earlier.
It turns out that in fact  power-like corrections in a
4D effective theory with such operator added by hand can
dominate over  genuine quantum effects 
 of similar form associated with compactification. 
This problem will be discussed in detail, together with the extent to
which power-like running can  be regarded as an
effect of the compactification alone.

The plan of the paper is as follows.
In Section~\ref{sec2} we review the results for the running of the gauge
couplings in 6D N=1 supersymmetric gauge theories on
orbifolds. We stress the role of the 
higher dimensional operator in the running of the gauge 
couplings at $k^2 v\sim \cO(1)$. In  Section~\ref{three}
a detailed numerical investigation  is presented for the running,
for momenta near the compactification scale(s). 
The Appendix contains  technical  formulae used in the text.

\section{Background field method: 
one-loop results and interpretation}\label{sec2}

We begin with a brief review of the results in \cite{DHK} for the 
momentum scale  dependence of the 4D effective gauge coupling, in 
4D N=1 supersymmetric (non-Abelian) gauge theories obtained from an
orbifold compactification. The starting point was a 
 general 6D N=1 gauge theory compactified on $T^2/Z_2$. 
The  associated action considered a hypermultiplet in representation $r$ of the 
gauge group $\cG$, with a  bulk gauge coupling $g$
and boundary conditions for the 6D gauge fields $A_M$ such as to 
ensure a 4D gauge theory (even $A_\mu$ and odd $A_{5,6}$,
$\mu=0,1,2,3$ with respect to $Z_2$ action). 
 A background field method was introduced which
considered quantum fluctuations of the gauge field about a classical background. 
Using this method, an  orbifold-compatible one-loop effective  action
for a classical background gauge field was computed in detail.
From this one can extract       the one-loop
self-energy of the  off-shell gauge boson.
For a full presentation of the effective action and detailed results 
the reader is referred to Section III of \cite{DHK}.
It was found that at one-loop, the 4D gauge field part $\Gamma^{(2)}[A_\mu]$ of 
the effective action can be written 
as a sum of  a bulk part and  a brane-localised part:
\be
\Gamma^{(2)}[A_\mu]=\Gamma_{\rm bulk}+\Gamma_{\rm brane}
\ee
$\Gamma_{\rm bulk}$ is the result of integrating all modes associated
with the compactification, while $\Gamma_{\rm brane}$ is generated
because of the localised singularities of the orbifold  $T^2/Z_2$ 
which break translation invariance and thus 
momentum conservation in the compact
directions.
 Their expressions are (see eqs.(3.1), (3.14), (3.49),
(3.50) in \cite{DHK}):

\bea
\label{gamma1}
\Gamma_{\rm bulk}&=&\frac{1}{2}\sum_{{\vec k},{\vec k}'}
\int \frac{d^4 k}{(2\pi)^4}
A^a_\mu(-k,-{\vec k}') A^a_\nu(k,{\vec k})
\big((k^2-{\vec k}^2)g^{\mu\nu}-k^\mu k^\nu\big) \nonumber \\
&&
\quad \times
\bigg[-\frac{1}{g^2}-i\Big(C_2(G)-C(r)\Big)\, 
\Pi^{\rm bulk}(k,\vec k')\bigg]\delta_{\vec k, \vec k'},
 \\[9pt]\Gamma_{\rm brane}
&=&
\!\frac{1}{2}\!\sum_{{\vec k},{\vec k}'}
\!\int\!\! \frac{d^4 k}{(2\pi)^4}
A^a_\mu(-k,-{\vec k}')A^a_\nu(k,{\vec k})
\big(k^2g^{\mu\nu}\!-\! k^\mu k^\nu\big)
\Big[\!-\!4iC_2(G)\Pi^{\rm local}(k,\vec k,\vec k')\Big]\qquad 
\label{gamma2}
\eea
Here $\vec k$ has two components $(n_1/R_5, n_2/R_6)$, that
 denote the discrete momenta in the two 
compact directions, with $n_1, n_2\in\bZ$ standing for the Kaluza-Klein
levels and $R_{5,6}$ the compactification radii. Further, $A_\mu(k,\vec k)$ is the 
gauge field, $g^{\mu\nu}$ is the 4D metric  $g^{\mu\nu}=diag(1,-1,-1,-1)$; $g$ 
is the 6D bulk gauge coupling (of mass dimension $-1$)  and
$C_2(G)$ and $C(r)$ are the Casimir
operators\footnote{
Casimir operators for a
representation $r$ denoted $G$ ($N$) for the adjoint
(fundamental) representation  are:
$ {\rm tr}(t^a_G t^b _G)=C_2(G)\delta_{ab}, \,
{\rm tr}(t^a_r t^b_r)=C(r)
\delta^{ab}$,
with $C_2(G)\!=\! C(G)\!=\!N$, $C(N)\!=\! 1/2$; $C_2(N)\!=\!(N^2-1)/2N$
for $SU(N)$} for adjoint and $r$-representation, respectively.
From (\ref{gamma1}), (\ref{gamma2}) one immediately identifies the scalar part
of the 4D gauge boson self-energy,  with $k^2$ ($k^2,\vec k^2$) 
denoting the 4D (6D) external  (momentum)$^2$ inflow in the associated
one-loop diagrams, respectively.

As indicated by the two Casimir operators present in its expression,
$\Gamma_{\rm bulk}$ is generated by both the vector multiplet
($C_2(G)$ dependence) and hypermultiplet ($C(r)$ dependence).
Unlike $\Gamma_{\rm bulk}$, $\Gamma_{\rm brane}$
receives contribution only from  the vector multiplet (only $C_2(G)$
dependence)  and corresponds
to a brane counterterm to gauge coupling. Finally,
$\Pi^{\rm bulk}$ and $\Pi^{\rm local}$ are
the bulk and brane-localised contributions to the effective action
and have the expressions
\smallskip
\bea
\Pi^{\rm bulk}(k,{\vec k}') &\equiv& 
 \mu^{4-d} \sum_{\vec p'}\int \frac{d^d p}{(2\pi)^d}
\frac{1}{(p^2-{\vec p}^{\prime 2})\,[(p+k)^2-(\vec p'+\vec k')^2]}, 
\label{pi-bulka} \\[9pt]
\Pi^{\rm local}(k,\vec k,\vec k')&\equiv & 
\frac{\mu^{4-d}}{2} \int
\frac{d^d p}{(2\pi)^d}\frac{1} %%% \delta_{-2{\vec p}',{\vec k}'-{\vec k}}
{(p^2-({\vec p}')^2)\,[(p+k)^2-({\vec p}'+{\vec k}')^2]}\bigg\vert_{\,\,\vec
  p'=(\vec k-\vec k')/2} \label{pi-g2} 
\eea

\medskip
\noindent
Here $d=4-\epsilon$ and $\mu$ is a finite mass scale introduced by dimensional
regularisation (DR). Below we are only interested in the
case $\vec k=\vec k'=0$ i.e. of zero-mode gauge coupling.
One shows \cite{DHK}:
\smallskip
\begin{eqnarray}
\Pi^{\rm local}\big(k,0,0\big)
&=&\frac{i}{32 \pi^2}\bigg\{\frac{2}{\epsilon}+\ln 4\pi e^{2-\gamma_E}
+\ln \frac{\mu^2}{k^2}\bigg\},
\nonumber\\[12pt]
\!\Pi^{\rm bulk}(k, \vec 0)
&=& 
\frac{i\, (2\pi \mu)^{\epsilon}}{(4\pi)^2\,v}
\int_0^1 dx \,\cJ\Big[x(1-x) k^2\Big]\sim \frac{k^2}{\epsilon},\qquad\quad 
v \equiv 4\pi^2 R_5 R_6.\label{qq1}
\eea

\medskip
\noindent
The expression of $\cJ$ and of its integral are 
presented in the Appendix in eqs.(\ref{M1(0)text}) to (\ref{jjj1}). 
$\cJ$ has a pole which is the result of summing in $\Pi^{\rm bulk}$  
individual  contributions from  infinitely many Kaluza-Klein modes. 
While each individual mode gives a pole $1/\epsilon$, after re-summing these 
divergent contributions \cite{DHK} one obtains a
single pole in $\Pi^{\rm bulk}$\footnote{This pole
  is the singularity  of  Hurwitz zeta function
 $\zeta[1+\epsilon,w]\!=\!1/\epsilon\!+\!\cO(\epsilon^0)$
entering in  the re-summed result}.
As shown in (\ref{qq1}), this pole depends on the  external momentum $k^2$; 
the gauge boson self energy (scalar part) is 
proportional to $k^2/\epsilon$ and a {\it higher derivative}
counterterm is then  needed. The poles in $\Pi^{\rm bulk}$ 
and in $\Pi^{\rm local}$ of (\ref{qq1}) are then cancelled by the
  following new terms  in the action:
\smallskip
\bea\label{cct}
\cL'\!=\int d^2 z \, d^2 \theta\,
\bigg[\frac{1}{2 h^2_{\rm tree}} {\rm Tr}\, W^\alpha \Box_6 W_\alpha
\!+\!\frac{1}{2}\sum_{i=1}^4\frac{1}{g^2_{{\rm brane}, i}}
 {\rm Tr} \,W^\alpha\,W_\alpha
 \delta^{(2)}(z\!-\!z_{0}^i)\bigg]
+ {\rm h.c.}
\eea

\medskip
\noindent
Here $z_{0}^i\,\,(i=1,\cdots,4)$  are the fixed points of the
$T^2/{Z}_2$ orbifold.
Further,  $h^2_{\rm tree}$ is the additional dimensionless bulk coupling of the
higher derivative operator (which  ``absorbs'' the pole
$k^2/\epsilon$); $g_{{\rm brane},i}$   is 
a dimensionless brane coupling at the fixed point $z_{0}^i$
(and it ``absorbs'' the pole $1/\epsilon$).
The 4D {\it effective} gauge coupling, hereafter denoted $g_{\rm eff}$,
can be identified as the overall coefficient present in:
\smallskip
\bea\label{tree}
-\frac{1}{2}\,\,
{\rm Tr}\bigg[F_{\mu\nu}\bigg(\frac{1}{g^2_{\rm tree}}
+\frac{v}{h^2_{\rm tree}}\Box_4\bigg)F^{\mu\nu}\bigg],\qquad
{\rm where }\qquad
\frac{1}{g^2_{\rm tree}}\equiv\frac{v}{ g^2}
+\sum_{i=1}^4 \frac{1}{g^2_{{\rm brane},i} }.
\eea

\medskip
\noindent
with $g^2$  the bulk tree-level  coupling.
 Bringing together all contributions to $g_{\rm eff}$, one has 
from eqs.(\ref{gamma1}) to (\ref{cct}):
\medskip
\bea\label{run0}
\frac{1}{g^2_{\rm eff}(k^2)}=\frac{1}{g^2_{\rm tree}}
-\frac{k^2 \,v}{ h^2_{\rm tree}}+ i
 \Big[C_2(G)-C(r)\Big]\,v\,
\,\Pi^{\rm bulk}(k,0)
+4 i\, C_2(G)\,\,\Pi^{\rm local}(k,0,0).\quad
\eea

\medskip
\noindent
This equation provides the most general expression for the
corrections due to  bulk and brane-localised terms, to the
4D effective gauge coupling $g_{\rm eff}(k^2)$.
It gives the running of  $g_{\rm eff}(k^2)$  which can be used for
any such orbifold  model with a specified gauge group and matter content.
 While the detailed expression of $\Pi^{\rm
  bulk}$   is complicated, in some limiting cases one  easily obtains
interesting formulae which describe the running of $g_{\rm eff}(k^2)$.

To study the running $g_{\rm eff}(k^2)$
we first introduce the following notation:
$u=R_6/R_5$, with $b_1=-3 C_2(G)+C(r)$ and $b_2=C_2(r)-C(G)$ 
for  the familiar N=1 and N=2
 one-loop  coefficients, respectively.
After some algebra one finds for
$k^2 \ll 1/R_{5,6}^2$:
\smallskip
\bea\label{run1}
\frac{4\pi}{g^2_{\rm eff}(k^2)}
=
\frac{4\pi}{g^2_{\rm tree}}
+\frac{b_1}{4\pi} \ln\frac{\xi_1\,\mu^2}{k^2}
-
\frac{b_2}{4\pi}\ln\bigg[4\pi\,e^{-\gamma_E} \mu^2\, v \, u\,
  \vert \eta (i u)\vert^4\bigg]
-\sigma
\eea

\medskip
\noindent
where $\xi_1= 4\pi e^{2-\gamma_E}$ and $\sigma=\cO(k^2 v)\ll 1$ is a small
correction (discussed below).   It also 
contains the overall threshold correction from Kaluza-Klein modes
represented by the term proportional to $b_2$, and
 which is momentum independent.
Finally, the term $\sigma$ is suppressed at low
$k^2 v\ll 1$ and is given by
\medskip
\bea\label{run2}
\sigma=- \frac{4\pi\,k^2\, v}{h^2_{\rm ren}}+\cO(k^4 v^2),\quad
{\rm where}\qquad
\frac{4\pi}{h^2_{\rm ren}}=
\frac{4\pi}{h^2_{\rm tree}}+\frac{b_2}{96 \pi^2}\,\ln \bigg[\pi\,
e^{\gamma_E}\,\mu^2  \,R_5^2\,\vert\eta(i\,u)\vert^{-4}\bigg]
\eea

\medskip
\noindent
Eq.(\ref{run1}) can be written at two different momentum scales, 
and the difference of the results obtained provides the {\it running} 
of $g_{\rm eff}$ wrt $k^2$  for $k^2\ll 1/R_{5,6}$. 
The  running is independent of
the overall threshold of the massive modes, and one recovers the usual
logarithmic running   (proportional to $b_1$), present
in 4D N=1 renormalisable theories such as the MSSM.
At such low momenta the massive modes are totally decoupled and 
do not affect the running of $g_{\rm eff}$ and the theory appears 
4D renormalisable.

If we increase $k^2$ to values $k^2\sim \cO(1/R_{5,6}^2)$ 
the correction $\sigma$ in
(\ref{run1}) becomes important and 
cannot be neglected anymore.
The running of the effective coupling $g_{\rm eff}(k^2)$  then
deviates  from the usual, low-energy  4D  logarithmic running. 
The analytical results are in this case less illuminating and the
running in this region  is discussed in detail in the next section.

There is another limit  derived from (\ref{run0}),
that of $k^2 \gg 1/R_{5,6}^2$, when $g_{\rm eff}(k^2)$ behaves like
\medskip
\bea\label{run3}
\frac{4\pi}{g^2_{\rm eff}\,(k^2)}
&=&\frac{4\pi}{g^2_{\rm tree}}
-\frac{2 \, C_2(G)}{4\pi} \ln\frac{\xi_1\,\mu^2}{k^2}
- \frac{4 \pi\,\,k^2\,v}{h^2_{\rm ren}(k^2)},
\eea
 where 
\bea\label{rrr4}
\frac{4\pi}{h^2_{\rm ren}(k^2)}&=&\frac{4\pi}{h^2_{\rm tree}}
+\frac{b_2}{96\pi^2}\ln\frac{\xi_2\,\mu^2}{k^2},
\eea

\medskip
\noindent
with $\xi_2=4\pi e^{8/3-\gamma_E}$ a scheme-dependent subtraction
constant. Therefore, if $k^2\gg 1/R_{5,6}^2$,  a linear (``power-like'') 
dependence on the  momentum scale $k^2$ dominates the running of
$g_{\rm eff}$, with a coefficient controlled by the coupling of the 
higher derivative operator $h_{\rm ren}$. Notice that $h_{\rm ren}$ has now a 
logarithmic dependence on $k^2$, unlike in (\ref{run2}). While the
running of $h_{\rm ren}$ is
subtraction scheme independent (i.e. no dependence on $\xi_2$),
 however, the  term $k^2/h_{\rm ren}^2(k^2)$  entering in the 
running of $g_{\rm eff}(k^2)$ is not. Such problem does not arise for 
$k^2\ll 1/R^2_{5,6}$ of eq. (\ref{run1}), since in that case any scheme 
dependence in $\sigma$ is entirely suppressed at low $k^2$.

Let us mention that the ``power-like'' running of gauge couplings
was  extensively considered   in the past, to  often mean 
a dependence of $g_{\rm eff}$ on  some high UV cutoff scale
$\Lambda$, rather\footnote{The relation between  one-loop 
results in DR and UV cutoff schemes was discussed in
 Section 5 of \cite{DHK}.} 
than on a momentum scale $k^2$  considered here. 
In such  picture it is not apparent however,
 why a  dependence of $g_{\rm eff}$ on 
$\Lambda R$ (instead of $k R$), which is momentum independent,
is suppressed and decoupled at low  momentum $k^2$, 
 when only logarithmic corrections should be present.
 Our results  show that, ultimately,
 the power-like  running of $g_{\rm eff}(k^2)$ 
is due to the higher derivative operator  generated dynamically at
one-loop.
This operator comes with a  coefficient which is indeed strongly suppressed
 at low  $k^2$ relative to the compactification scale(s). In this limit higher 
derivative operators indeed decouple, to restore  renormalisability of the
low-energy  4D theory and leave only the logarithmic running,
as it should be the case.

While the limits   $k^2\ll 1/R_{5,6}^2$ and $k^2\gg 1/R_{5,6}^2$
discussed so far  provide simple analytical formulae eqs.(\ref{run1})
to (\ref{rrr4}) with a clear behaviour,
other cases of interest such as $k^2\sim 1/R^2_{5,6}$,
can only  be studied numerically.  This region is important  for 
experimental searches for signatures  associated with large 
radii of compactification (TeV-scale region), 
 it was not investigated in the past and
is discussed  in detail in the following.

\section{One-loop running across the compactification scales.}
\label{three}

We  address here the running of the 4D  coupling  $g_{\rm eff}(k^2)$
for  momentum scales near the compactification scale(s) 
$k^2\!\sim \!{1}/{R^2_{5,6}}$  using a  numerical  
study based on eqs(\ref{qq1}), (\ref{run0}), (\ref{M1(0)text}) to (\ref{jjj1}).

One question we  address is to what extent a deviation 
of the running of the 4D  coupling  from the 
logarithmic one (specific to SM and MSSM) would be  an indication of new
physics associated with quantum corrections due to ``large'' extra dimensions.
Contrary to a widespread opinion,  we argue that in general the 
observation of a linear (``power-like'') dependence of the 
gauge couplings on $k^2$ as opposed to the logarithmic one is not
necessarily an indication of the existence of extra dimensions. 
Such dependence can simply be ``obscured''
 by classical physics, represented by a  higher derivative
 operator, which can always be present in  a 4D effective field
 theory and which is {\it not} derived from  compactification of a higher 
dimensional theory.

Further, regardless of the origin of the higher derivative operator 
(classical or dynamically generated by the bulk), its coefficient
is in both cases unknown. 
In the classical case of adding the operator in the 
tree-level action, allowed on symmetry grounds, its
coefficient is an unknown parameter of that  4D effective theory.
If  the higher derivative
operator is however dynamically generated by compactification,
at one-loop level, one can argue that its coupling should be
smaller than that of the classical case, and one should be able to
distinguish this case from the ``classical'' scenario.
  However, this is not the whole
story; there are further  
uncertainties due to the subtraction scheme used in the process
of the renormalisation of the higher derivative operator (see later).
To conclude,  in order to fully understand the {\it quantum} behaviour of 
 the 4D coupling due to extra dimensions, one needs to
 know a new parameter, this time the coupling  
of the higher derivative operator.
Quantum consistency of the calculation
 requires the introduction of new (classical)
parameters in the theory, particularly relevant for scales near $1/R_{5,6}$. 
This  is common in non-renormalisable theories and
is expected to  happen again  in higher loops.

So can then one indeed distinguish  quantum effects to $g_{eff}$ associated with 
compactification, from classical higher derivative operators
 added explicitly in a  4D N=1  effective theory\footnote{not 
derived from a compactification of a higher
  dimensional theory.}? To address this, 
consider the most general form of the running of the effective 
gauge coupling;  from (\ref{run0}) one can write,
 with $h_{\rm ren}$ of eq.(\ref{run2})
\smallskip
\bea\label{RR}
\frac{4\pi}{g^2_{\rm eff}(k^2)}
=
\frac{4\pi}{g^2_{\rm tree}}
+
\frac{b_1}{4\pi} \ln\frac{\xi_1 \mu^2}{k^2}
-\frac{b_2}{4\pi}\ln\Big[4\pi e^{-\gamma_E} 
\mu^2\, v \, u\,
\vert \eta (i u)\vert^4\Big]
-  \frac{4\pi\,k^2\, v}{h^2_{\rm ren}}
+\,\frac{b_2}{4\pi}\,\,\cZ[k^2]
\eea

\medskip
\noindent
The quantity  $\cZ[k^2]$ contains terms of order $(k^2 v)^n$ with
 $n>1$ and can be neglected if $k^2 v\ll 1$, since it is
 sub-leading to the term in front of it;  at low
scales  one recovers eqs.(\ref{run1}), (\ref{run2}).
The exact expression of $\cZ[k^2]$ in terms of the function $\cJ$ 
is  (see Appendix, eq.(\ref{rr})):

\bea  \label{text}
\cZ[k^2]\!=\!\!\int_0^1\!\! dx\,\cJ[x(1-x) k^2]\!-\!
\frac{\pi\,k^2}{6} R_5 R_6 \bigg[\frac{-2}{\epsilon}
\!+\!\ln\frac{4\pi e^{-\gamma_E} \vert \eta(i u)\vert^4}{4
  \pi^2 \mu^2 R_5^2}\bigg]
\!+\!
\ln \bigg[4\pi^2 e^{-2} \vert \eta(i u)\vert^4 k^2 R_6^2\bigg]
\eea

\medskip
\noindent
where $u=R_6/R_5$ and the pole $k^2/\epsilon$ is cancelled by
that in the integral of $\cJ$, to give a finite $\cZ$ (see also
 eqs.(\ref{M1(0)text}), (\ref{M1(0)stext}), (\ref{jj1}), (\ref{jjj1})).
Writing eq.(\ref{RR}) at two different momentum scale
one finds  for the running of $g_{\rm eff}(k^2)$  
\medskip
\bea\label{RRR}
\frac{4\pi}{g^2_{\rm eff}(k^2)}=\frac{4\pi}{g^2_{\rm eff}(q^2)}
+
\frac{b_1}{4\pi} \ln\frac{q^2}{k^2}
-  
\frac{4\pi\, v}{h^2_{\rm ren}}(k^2-q^2)
+\frac{b_2}{4\pi}\,\,\Big[\cZ[k^2]-\cZ[q^2]\Big]
\eea

\medskip
\noindent
This is the main result of the paper, which is
important for phenomenological applications
which try to identify quantum effects of extra dimensions. It can be
easily applied to specific models (i.e. specific gauge group and
matter spectrum such as that of the MSSM) and is used in the
analysis below.

The term $k^2 v/h^2_{\rm ren}$ in $g_{\rm eff}(k^2)$ 
was  introduced as a counterterm and its 
coefficient $h_{\rm ren}^2$ is subtraction scheme dependent.
As argued earlier, this term can also have a classical origin 
in the 4D  effective field theory, again with a model-dependent
coefficient. Thus, one  cannot identify its exact origin.
However, any deviation from this contribution in the 
running in (\ref{RR}) and (\ref{RRR}), represents a correction from
the massive bulk modes - other than
 that proportional to $k^2$ - which cannot be 
described by a tree level D=6 higher derivative operator or by 
$\ln k^2$. This correction is due to 
$\cZ(k^2)$ alone, is specific to  compactification and is
investigated  below.

From  eq.(\ref{RRR}) one  also finds 
the one-loop effective beta function, which depends on the momentum scale:
\medskip
\bea\label{beta}
\beta[k^2]\equiv \frac{4 \pi  d g^{-2}_{\rm eff}(k^2)}{d\ln k^2}=
-\frac{b_1}{4\pi}-\frac{4\pi}{h^2_{\rm ren}}\,\,v\,k^2
+
\frac{b_2}{4\pi}\,\frac{d}{d\ln k^2}\, \,\cZ[k^2]
\eea

\medskip
\noindent
with $h_{\rm ren}$   given by (\ref{run2}), which is momentum
independent\footnote{We use a separation in (\ref{beta}) with
$h_{\rm ren}$   given by  eq.(\ref{run2}) and not by 
 eq.(\ref{rrr4}) because this ensures a momentum independent
coefficient of the $k^2$-like term, induced by 
a higher derivative operator, as in the ``classical''
case, when added to the theory at the
tree level. The dependence of $h_{\rm ren}$ on the momentum as in 
 eq.(\ref{rrr4}) is a compactification effect  included, with other
 similar terms,  in $\cZ$.}.
The first term in (\ref{beta}) is the usual coefficient of the beta function
 in 4D N=1 models, and the only present when $k^2\ll 1/v$.
The second term is a leading  correction  of
compactification for $k^2 v< 1$, and brings  a momentum dependence to
the beta function;  this term can  also be present, together with the first term,
in any 4D N=1 effective model with a higher derivative operator added 
{\it explicitly} to the action  with a coefficient $\zeta=4\pi v/h_{\rm ren}^2$.
Additional corrections to $\beta[k^2]$, represented by the derivative of $\cZ$ 
 involve higher orders in  
complicated series of powers of $k^2 v\ll 1$ and cannot be accounted 
for by  ``classical'' higher derivative  operators,
in the approximation considered here.
  These corrections  are important for $k^2 v\sim 1$ and are
investigated below.

For simplicity, we consider that the two
compactification scales are equal $R_5=R_6=R$, to  reduce the 
number of parameters of the study. We  analyse the dependence of
$\cZ[k^2]$ and of its derivative $d(\cZ[k^2])/d (ln k^2)$ as we 
vary $k^2$ 
through  values  that cross the compactification scale~$1/R^2$~-~this
can be referred to as  the ``crossover'' region.

In  Figure~\ref{fig1},   $\cZ[k^2]$ and its logarithmic derivative are
 plotted in function
of  $\log_{10}(k^2 R^2)$. For $k^2 \!\ll\! 1/R^2$, $\cZ$
 vanishes, and this shows the decoupling of the massive
bulk modes from the running of the coupling $g_{\rm eff}(k^2)$.
   As $k^2$ approaches $1/R^2$, $\cZ[k^2]$
 starts to increase, somewhat surprisingly,
 {\it before} reaching the scale $1/R^2$! 
This is an interesting effect, which 
suggests that  we sample  global effects  of an {\it infinite} set of
Kaluza-Klein states rather than individual Kaluza-Klein modes; 
such effects did  not   decouple at a 
{\it fixed} threshold ($1/R$) as one would naively
   expect\footnote{A decoupling is not necessarily present 
in the case of including the effect of an infinite set/tower of states.};
instead, the infinite set 
of Kaluza-Klein states gave a remnant effect even below this
scale. This 
 effect is associated with the infinite series expansion present in $\cJ_<$ in
 (\ref{M1(0)text}), see also (\ref{jj1}), (\ref{rr}), which gives
  significant effects close to the compactification scale.

Another interesting effect,  as seen in 
both plots of $\cZ[k^2]$, is that there exists a smooth transition 
while crossing the compactification scale threshold, and also the 
derivative of the coupling (in fact the last term in $\beta[k^2]$) is continuous. 
At the technical level, note  that
the plots use very different series expansions of the bulk correction, 
below and above the  scale\footnote{See the explicit formulae in the
appendix, $\cJ_<$, $\cJ_>$ and $\cZ$ of eqs.(\ref{M1(0)text}) to
(\ref{rr}).} $k^2=4/R^2$. The  behaviour of $\cZ[k^2]$ and 
thus of $g_{\rm eff}(k^2)$
and $\beta[k^2]$  while crossing the
compactification scale together with the 
presence of smooth, power-like dependence even
below $1/R^2$, suggest that one is not be able to see from the
running alone, {\it individual} effects of Kaluza-Klein states\footnote{
which when decoupling  at a fix scale would spoil the smooth
behaviour we see.}, but an overall effect of all modes.
We only see  a smooth transition (crossover) to a higher dimensional
theory, with the massive modes giving a  global effect to the running.

\begin{figure}[t]
\begin{tabular}{cc|cr|}
\,\,\parbox{7.8cm}{\psfig{figure=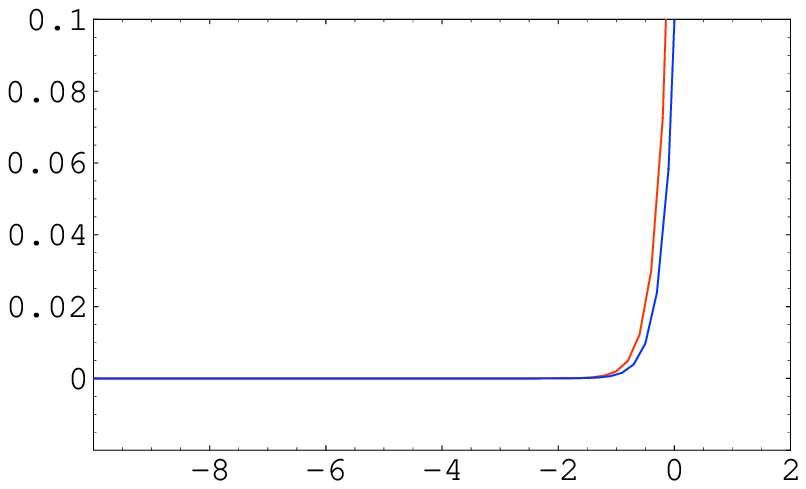,height=4.2cm,width=7cm}}
\hfill{\,\,}\
\parbox{7.8cm}{\psfig{figure=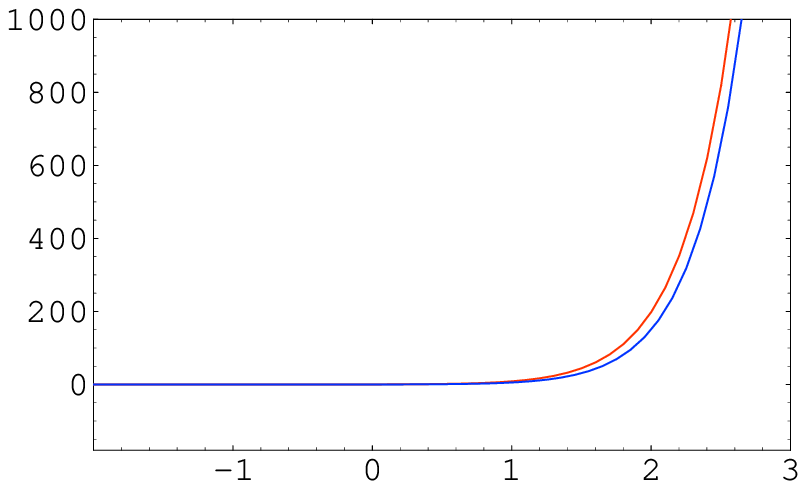,height=4.2cm,width=7cm}}
\end{tabular}
\def\baselinestretch{1.1}
\caption{\small 
The dependence of  $\cZ[k^2]$ on the momentum scale (blue/right curves)
and of its contribution $d \cZ[k^2]/d(\ln k^2)$ to the beta function 
(red/left curves).
The horizontal axis corresponds to $\log_{10}(k^2 R^2)$ in both graphs.
At low $k^2$,  $\cZ\ra 0$, while for $k^2\sim 0.1/R^2$ or larger
a  non-vanishing correction  sets in due to massive (bulk) modes. Their
one-loop effects are thus manifest even below $1/R^2$.
There is a smooth transition while $k^2$ 
is crossing  the threshold of compactification
scale $1/R^2$, for both $g_{\rm eff}^2(k^2)$ and $\beta[k^2]$.}
\label{fig1}
\end{figure}

\begin{figure}[t]
\begin{tabular}{cc|cr|}
\parbox{8.cm}{\psfig{figure=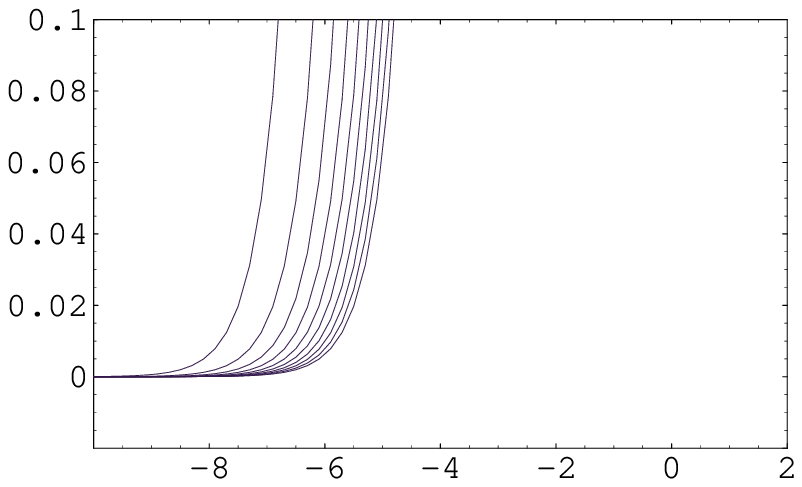,height=4.2cm,width=7.3cm}}
\hspace{-0.5cm}
\parbox{8.cm}{\psfig{figure=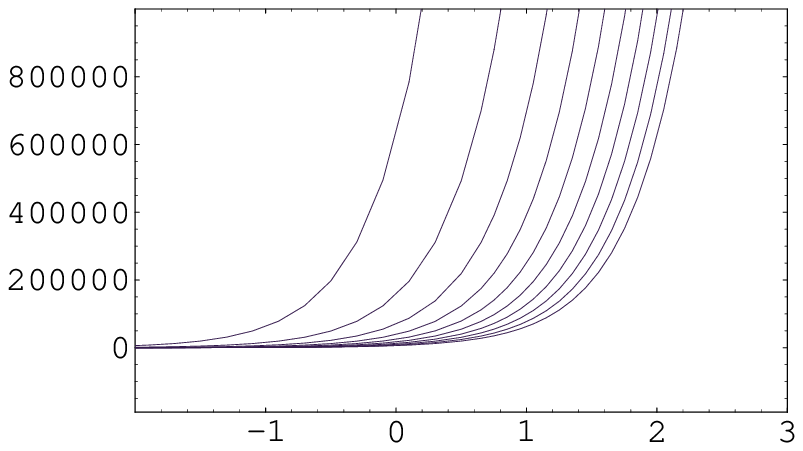,height=4.4cm,width=7.7cm}}
\end{tabular}
\def\baselinestretch{1.1}
\caption{\small 
The dependence of the total correction 
represented by $\cZ[k^2]-(4\pi)^2 k^2 v/(b_2 \,h^2_{\rm ren})$ which thus
includes ``classical'' effects as well,
assuming a range of values from $h_{\rm ren}^2\,b_2=-0.1$ (leftmost curve)
to $-1.$  (the rightmost curve) with step $0.1$. These results
are much larger than those in Fig.\ref{fig1} for similar range of
values for $\log_{10}(k^2 R^2)$. In fact $\cZ[k^2]$ is totally
negligible compared to $-(4\pi)^2 k^2 v/(b_2 \,h^2_{\rm ren})$
for the quoted range of values for $b_2\,h_{\rm ren}^2$. The derivative
with respect to $\ln k^2$ of the above expression entering
$\beta[k^2]$ has identical
values due to the $k^2$-like behaviour. Larger values
(in absolute value)
for $b_2 \,h_{\rm ren}^2$ can in principle exist. For 
$b_2\,h_{\rm ren}^2$ of opposite sign, these plots are reflected
with respect to the $OX$ axis.}
\label{fig2}
\end{figure}

Note  that the part of the effective beta
function represented by $d\cZ[k^2]/d(\ln k^2)$ and seen in 
Figure~\ref{fig1} has within a very good approximation, an
identical behaviour to $\cZ$. This suggests that the momentum
dependence of the coupling due to  $\cZ$ eq.(\ref{RR}), (\ref{RRR}),
remains almost power-like
(i.e. linear in $k^2 R^2$) for all $k^2$ in the range of 
non-vanishing  $\cZ$. This is an interesting result since such behaviour is 
the result of summing  series with a complicated 
dependence  on the momentum, as already mentioned
(compare $\cJ_<,\cJ_>$ of 
(\ref{M1(0)text}), (\ref{M1(0)stext}), (\ref{jj1}) to (\ref{rr})).
This ends our
discussion of pure quantum effects near the compactification scale,
that can be identified as due solely 
to  compactification of extra dimensions.

The next step is to compare the above quantum effects due to extra
dimensions  to other corrections present in $g^2_{\rm eff}$ of eqs.(\ref{RR}),
(\ref{RRR}), such as the $ k^2 v/h^2_{\rm ren}$ term.
To examine this term
relative to the $\cZ$ correction in Figure~\ref{fig1},
we present in Figure~\ref{fig2} 
 the expression $\cZ[k^2]-(4\pi)^2 k^2
v/(b_2 \,h^2_{\rm ren})$ entering in $g^2_{\rm eff}$,
 for $b_2 \,h_{\rm ren}^2 <0$.
From these figures it is clear that the relative
effect of $\cZ$ is actually small and can be
neglected. 
Further, the logarithmic derivative of this expression 
controls $\beta[k^2]$ and has an identical behaviour
 because
in Fig.~\ref{fig2} the value of $\cZ$ is small compared to 
 $(4\pi)^2 k^2 v/(b_2 \,h^2_{\rm ren})$ for the values of
$h_{\rm ren}^2 b_2$ considered (similar considerations for
 $b_2 \,h_{\rm ren}^2 >0$).  It is possible that 
$\cZ[k^2]$ and $(4\pi)^2 k^2 v/(b_2 \,h^2_{\rm ren})$
 reach comparable values, but this 
only happens at  large values of $h_{\rm ren}^2 b_2\sim 70$. In such cases 
$\cZ$ can dominate the power-like  term, but 
this region may reside outside perturbation 
theory in $h_{\rm ren}$ and is not  considered here.

Comparing Figures~\ref{fig1} and \ref{fig2}
for similar range of values for $k^2 R^2$, it is obvious that the
term proportional to $k^2 v$  dominates the effects
due to $\cZ$ of Figure~\ref{fig1}. Further,  from Figure~\ref{fig2}
one is unable to decide whether the linear dependence on $k^2$ is 
due to a higher derivative operator
 added by hand in  a 4D N=1 effective model 
but with a suitable coefficient, or due  
to compactification (i.e. $k^2 v/h^2_{\rm ren}$). 
To conclude, $k^2$-like terms in general dominate,
but their coefficient is model dependent (former case) 
or subtraction scheme dependent (latter case). Therefore,
identifying a $k^2$-like behaviour is not necessarily a signature
of quantum effects due to extra-dimensions, 
as often thought.

How generic are these findings for the running of the 
4D effective
gauge coupling in gauge theories on orbifolds?
To see this, recall that our results are a consequence of the presence of a
higher derivative operator and its origin can be: 
1) this operator is added explicitly in any 4D effective field
theory, and suppressed by some high scale;
or 2) this operator is generated  radiatively  as counterterm
to the gauge boson self-energy, as in
our 6D N=1 model compactified on $T_2/Z_2$. 
In the last case, the presence of the 
higher derivative operator is
ultimately  related to 
the number of Kaluza-Klein (KK) sums  present in the loop correction,
which triggered a divergence when summing all individual modes'
effects\footnote{
This divergence is also related to the fact that the
classical propagator in two dimensions is divergent; however
this does not apply to 5D orbifolds.}.
The effects we found can therefore
be present also in  5D orbifolds beyond one-loop order, 
when more KK sums are present  \cite{santamaria}.
Ultimately, it is the non-renormalisability of the
higher dimensional theory which brings back the presence of such
operators at the quantum level. They are therefore a common
presence, dynamically generated
in higher dimensional theories compactified to 4D 
\cite{nibbelink}-\cite{santamaria}.

One could ask whether the above results for the running of the
4D effective gauge couplings could affect the gauge couplings unification in 
 higher dimensional models that can  recover, upon
 compactification, the Minimal Supersymmetric Standard Model (MSSM).
The answer is negative for the following reason, often overlooked.
In general the test of unification of
 couplings is not necessarily a test of the {\it running} 
itself at the high
scale, but rather a test of the compatibility of the low energy 
measurements of the couplings with the existence of a 
 gauge group independent coupling at some high scale,
{\it after} integrating out all massive modes and other states.
 In this situation, all higher dimensional  operators 
whose effects we discussed above are
suppressed  at 
low $k^2\sim M_Z^2$, as  seen explicitly in
eq.(\ref{run1}).
According to this equation, the renormalisation scale $\mu$ introduced
by dimensional regularisation plays the role of unification scale.
The only power-like dependence of the couplings with respect to any
scale or momenta is present in $\sigma$ and this is strongly suppressed
when $k^2\ll 1/v$, i.e. at low scales. Also any scheme dependence
for the subtraction of the divergence of the higher derivative
operator (present in $\sigma$) is also strongly suppressed. Such
suppression is important, since it ensures   a meaningful (i.e. scheme independent)
analysis of unification in  higher dimensional models. Finally, the only part of 
eq.(\ref{run1}) which ``feels'' the bulk effects is the
overall threshold of integrated out massive states (proportional to $b_2$
and momentum scale independent). Note that this threshold
is similar to that  obtained in string calculations at 
one-loop \cite{RC5}\footnote{This
  agreement is not too surprising - in both cases there are no winding modes
corrections to the threshold}. For a discussion of the
threshold corrections on unification in 6D orbifold models of type discussed 
here see \cite{HML}.

\section{Conclusions}

We investigated in detail the dependence on the {\it momentum}
scale  of the 4D effective gauge coupling  and of its beta function 
in field theory orbifold compactifications of 6D N=1 supersymmetric
theories on $T^2/Z_2$.
 Such dependence cannot  be investigated by on-shell string
 calculations   and this stresses the importance of our
 results.   Our purpose was to  study in detail
 the one-loop behaviour of the 4D effective gauge coupling and its
beta function for values of  the momentum scale
{\it near} the compactification scale(s), where analytical 
formulae are  complicated and slowly convergent, 
and a numerical approach is required.
Near this scale, the results are controlled by the presence of a
higher derivative gauge kinetic term which in 6D gauge theories on orbifolds
is dynamically generated.

The results show that the running of the 4D effective gauge coupling 
remains a continuous function of the momentum,
 with a continuous derivative across
 the compactification scale(s)
$k^2\!\sim\! 1/R^2$.
The  smooth behaviour of the effective gauge coupling and of
its  beta function while crossing the compactification scale 
is an interesting effect.  We also showed that, rather 
surprisingly, the power-like  behaviour due to quantum effects of
extra dimensions sets in at momenta
 {\it smaller} than the compactification scale.
This  is a direct result of including 
the one-loop threshold effects of infinitely many Kaluza-Klein modes. 
The results showed
that there is a  smooth crossover  from a 4D theory to a higher 
dimensional theory, with the massive modes giving together an overall
{\it  global}  rather  than an individual effect to the running. 
Our formulae could be used in experimental searches for quantum effects 
of large extra dimensions.

Throughout the  ``crossover'' region, the running of the effective 
coupling is dominated by the power-like (linear) terms in the momentum
scale $k^2$ and this behaviour of the gauge coupling is controlled by 
the coupling of the higher derivative operator. Unfortunately, such
running cannot  immediately be
attributed to the existence  of extra dimensions,
as usually thought. This is because such running can
be also present  in any 4D effective field
theory with an action to  which one adds {\it explicitly} a  higher dimensional
(derivative) operator with a suitable coefficient.
This makes it very difficult 
to disentangle genuine quantum effects associated
with compactification, from tree-level, ``classical'' running of the
coupling.
However, after carefully isolating the power-like correction due to
 the higher derivative
operator itself, we identified additional effects 
 which are specific to the quantum effects of compactification only, in 
the approximation  considered.
Unfortunately, these effects (accounted for by $\cZ$) are in general
small compared to the leading term $k^2 v/h_{\rm ren}^2$ for $k^2\sim
1/v$, but become more important at large values of  $h_{\rm ren}^2 b_2$.

To conclude, identifying a power-like running of the effective gauge
coupling is not necessarily a  footprint of compactification and a
sign of existence of extra dimensions.
With a suitable coupling, a tree-level higher derivative operator
added to a 4D effective theory can  bring a running 
similar  to that  associated with compactification.
On the phenomenology side, these results show that
it  is difficult to decide, from the
running of the 4D effective coupling alone, when quantum  effects due to
extra dimensions are turned on; to this purpose
 additional effects such as direct production
of Kaluza-Klein modes etc, will be required.

\newpage
\section*{Acknowledgements:}
This work was supported partially by the RTN European Program
MRTN-CT-2004-503369, ``The Quest for Unification'' and by
the Marie Curie Research Training Network
of the European Community, contract n. MRTN-CT-2006-035863.

\bigskip
\section*{Appendix}
\def\theequation{A-\arabic{equation}} 
\setcounter{equation}{0}

{\bf [a].}  If $0\leq c/a_1\leq 1$, the  function $J$  of eq.(\ref{qq1})
is denoted $J_<$ and has the  expression

\begin{eqnarray}\label{M1(0)text}
&&
\cJ_{<}[c] 
=  \! \frac{\pi c}{\sqrt{a_1 a_2}} \bigg[\frac{-2}{\epsilon}\!+\!
\ln\Big[4\pi \,a_1\, e^{-\gamma_E}\Big]\bigg] 
-\sum_{n_1\in\bZ}\ln\Big\vert 1- e^{-2\pi \,\gamma (n_1)}\Big\vert^2
+
\frac{\pi}{3}\sqrt{\frac{a_1}{a_2}}-2\pi\sqrt{\frac{c}{a_2}}\qquad
\nonumber\\[6pt]
&&\qquad\qquad\qquad\qquad\qquad\qquad\qquad
- 2\,\frac{c\, \pi^\frac{1}{2}}{\sqrt{a_1 a_2}}\,
 \sum_{p\geq 1}\,\frac{\Gamma[p\!+\!1/2]}{(p\!+\!1)!}
\bigg[\frac{-c}{a_1}\bigg]^{p} 
\zeta[2p+1]
\end{eqnarray}

\medskip
\noindent
which is convergent under the above assumption. 
The following notation was used:

 $$\gamma(n_1)=(c+a_1\,n_1^2)^{1/2}/\sqrt{a_2},\,\,\,\,\,\,\, a_1=1/R_5^2,\,\,
\,\,a_2=1/R_6^2,\,\,\,\,\gamma_E=0.577216..$$

\medskip
\noindent
 and $\zeta[x]$ is
  the Riemann Zeta function.
Corrections of order $\cO(c)$ (recall that $c\sim k^2$ in the text), 
 come from the first square bracket  in $\cJ_<$  and also 
from the term involving $\gamma(n_1)$ which also brings in  a correction of type $\ln \,c$.
The remaining terms bring  sub-leading contributions of order
$c^2\propto k^4$ or higher.

\bigskip
\noindent
{\bf [b].} If $c/a_1>1$,  then $J$ of eq.(\ref{qq1})
 is denoted $\cJ_>$ and equals
\begin{eqnarray}\label{M1(0)stext}
\cJ_>[c]\!\!\!
 &=&\!\! \!
\frac{\pi c}{\sqrt{a_1 a_2}}\bigg[\! \frac{-2}{\epsilon}\!+\!
\ln\Big[
\!\pi \,c \,e^{\gamma_E-1}\Big]\!\bigg] 
\! -\!\sum_{n_1\in\bZ}\!\!\ln\Big\vert
 1\!-\!e^{-2\pi \,\gamma (n_1)}\Big\vert^2\!+\!
4\sqrt{\frac{c}{a_2}}\sum_{\tilde n_1>0}
\frac{K_1(2 \pi \tilde n_1 \sqrt{\frac{c}{a_1}})}{\tilde n_1}\nonumber
\\[-10pt]
\end{eqnarray}

\medskip
\noindent
where $K_1$ is the modified Bessel function which for large argument 
is exponentially suppressed:

\begin{eqnarray}
K_1[x]&=& e^{-x}\sqrt{\frac{\pi}{2
    x}}\left[1+\frac{3}{8 x}-\frac{15}{128 x^2}
+\cO(1/x^3)\right]\label{K_1}
\end{eqnarray}
The leading correction in $c\sim k^2$ to $\cJ_>$ comes again
from the linear term in $c$, while the two series have exponentially
suppressed terms. The pole structure is the same for both $\cJ_<$ and $\cJ_>$.

\medskip\bigskip
\noindent
{\bf [c].}
The integral present in $\Pi^{\rm bulk}$ of eq.(\ref{qq1})
is then,  if $k^2<4 a_1$:
\bea\label{jj1}
\int_0^1 dx \,\,\cJ[x(1-x)\,k^2]=
\int_0^{1} dx\,\cJ_<[x(1-x) k^2]
\eea
while if  $k^2>4 a_1$ then:
\bea\label{jjj1}
\int_0^1 dx \,\,\cJ[x(1-x)\,k^2]=
2\int_0^{f_k} dx\,\,\cJ_<[x(1-x) k^2]
+
2\int_{f_k}^{1/2} dx\,\,\cJ_>[x(1-x) k^2]
\eea
where $f_k=\frac{1}{2}\,\big[1-\big(1-4 a_1/k^2\big)^{1/2}\big]$.
For simplicity we assume $a_1=a_2=1/R^2$.

\bigskip\medskip
\noindent
{\bf [d].} The definition of the function $\cZ[k^2]$ used in the text
\bigskip
\smallskip
\bea  \label{rr}
\cZ[k^2]\!=\!\!\int_0^1\!\! dx\,\cJ[x(1-x) k^2]-
\frac{\pi\,k^2}{6 \sqrt{a_1 a_2}} \bigg[\frac{-2}{\epsilon}
+\ln\frac{4\pi a_1 e^{-\gamma_E} \vert \eta(i  u)\vert^4}{4 \pi^2 \mu^2}\bigg]
\!\!+\!\ln \bigg[4\pi^2 e^{-2} \vert \eta(i u)\vert^4 \frac{k^2}{a_2}\bigg]
\eea

\medskip
\noindent
where $u=\sqrt{a_1/a_2}$.
For small $k^2$, and with $a_1=a_2=1/R^2$,  $\cZ[k^2]\sim k^4 R^4$.

To evaluate the derivative of $\cZ[k^2]$ used in studying $\beta[k^2]$
of eq.(\ref{beta})
 one needs the derivative of
the integral of $\cJ$. This  is easily found from  (\ref{jj1}) if 
$k^2\leq 4a_1$, while if $k^2>4 a_1$ it is given by
\medskip
\bea
\frac{d}{d\,\ln \,k^2}
\int_0^{1}\!
\!\!\! dx \,J[x(1-x)\,k^2]=
2 \,f'_k \bigg[ \cJ_<[a_1] -\cJ_>[a_1]\bigg]
+
2\int_0^{f_k}\!\!\!
 dx \,\frac{\partial \cJ_<}{\partial \ln\,k^2}
+
2\int_{f_k}^{1/2}\!\! \!
dx\,\frac{\partial \cJ_>}{\partial \ln\,k^2}\,\,\,
\eea

\medskip
\noindent
where the argument of $J_{<,>}$ under the last two integrals is
$x(1-x) k^2$. 
It can be  checked that $\cJ_<[a_1]=\cJ_>[a_1]$ i.e.
the function $J$ is continuous so that
\medskip
\bea
\frac{d}{d\,\ln \,k^2}
\int_0^{1}\!
\!\!\! dx \,J[x(1-x)\,k^2]=
2\int_0^{f_k}\!\!\!
 dx \,\frac{\partial \cJ_<[x(1-x) k^2]}{\partial \ln\,k^2}
+
2\int_{f_k}^{1/2}\!\! \!
dx\,\frac{\partial \cJ_>[x(1-x) k^2]}{\partial \ln\,k^2}\,\,\,
\eea

\medskip
\noindent
This result was used in the text to
 compute $\beta[k^2]$ for $k^2\geq 4/R^2$.

\end{document}